\documentclass[aps,prd,reprint,twocolumn,groupedaddress,superscriptaddress]{revtex4-2}
\usepackage{todonotes}
\usepackage{dcolumn}
\usepackage{graphicx}
\usepackage{float}
\usepackage{physics}
\usepackage[colorlinks=true,allcolors=blue]{hyperref}
\usepackage{epstopdf}
\usepackage{changes}
\usepackage{mathtools}
\usepackage{tabularx} 
\usepackage{xcolor}
\usepackage{booktabs}
\usepackage{longtable}
\usepackage{array}
\usepackage{fancyhdr}
\usepackage{enumitem}
\usepackage{amsmath, amssymb}
\usepackage{amsthm} 
\usepackage{makecell}
\usepackage{hyperref}
\usepackage{bm} 
\usepackage{siunitx}

\begin{document}
\title{Discovering the Gell-Mann–Okubo Formula with Kolmogorov-Arnold Networks}

\author{Jian-Yao He}
\email{yaoyaoerg5251@163.com}
\affiliation{School of Nuclear Science and Technology, University of South China Hengyang, No 28, West Changsheng Road, Hengyang City, Hunan Province, China.}

\author{Xun Chen}
\email{chenxun@usc.edu.cn}
\affiliation{School of Nuclear Science and Technology, University of South China Hengyang, No 28, West Changsheng Road, Hengyang City, Hunan Province, China.}
\affiliation{INFN -- Istituto Nazionale di Fisica Nucleare -- Sezione di Bari, Via Orabona 4, 70125 Bari, Italy}

\author{Xiao-Yan Zhu}
\email{xyzhu0128@163.com}
\affiliation{School of Mathematics and Physics, University of South China, No. 28, West Changsheng Road, Hengyang City, Hunan Province, China.}

\author{Wen Luo}
\email{wenluo-ok@163.com}
\affiliation{School of Nuclear Science and Technology, University of South China Hengyang, No 28, West Changsheng Road, Hengyang City, Hunan Province, China.}

\begin{abstract}
Uncovering physical laws from experimental data is a fundamental goal of theoretical physics. In this work, we apply the spline-based, interpretable Kolmogorov–Arnold Network (KAN) to explore the algebraic structure underlying the baryon octet and decuplet mass spectra. Within a symbolic regression framework and without imposing theoretical priors, KAN recovers an equivalent symbolic representation of the classical Gell-Mann–Okubo mass relations, which can be further transformed into the standard form through algebraic rearrangement. Compared to conventional fitting approaches, this method achieves comparable predictive accuracy while offering substantially improved interpretability and analytic transparency. Our results demonstrate the potential of KAN as a powerful tool for symbolic discovery in hadron physics and for bridging data-driven modeling with fundamental physical laws.

\end{abstract} 

\maketitle
{\it Introduction:} 
Within the Standard Model, hadrons—composite particles made of quarks bound by the strong interaction—are described by Quantum ChromoDynamics (QCD). As a non‑Abelian gauge theory, QCD successfully accounts for phenomena such as color confinement and asymptotic freedom   \cite{gross1973ultraviolet,politzer1973reliable,callan1970broken,symanzik1970small,vanyashin1965vacuum,khriplovich1969green,fritzsch1973advantages,gross1973asymptotically,politzer1974asymptotic}, yet its non‑perturbative regime, which dominates the hadron spectrum, remains analytically challenging. Even so, the hadron mass spectrum exhibits patterns that reflect the underlying SU(3) flavor symmetry \cite{gell2018eightfold,okubo1962note,gell2018symmetries}. In the 1960s, Gell‑Mann and Ne’eman introduced the Eightfold Way classification, organizing the rapidly expanding set of experimentally discovered hadrons into octets and decuplets according to symmetry principles \cite{gell2018eightfold,okubo1962note,gell2018symmetries}. This framework indirectly led to the quark model, in which baryons and mesons are understood as three‑quark states and quark–antiquark pairs, respectively. The first observation of the exotic candidate X(3872) in 2003 challenged this simple three-quark picture. Since then, numerous exotic candidates have been reported at major high‑energy accelerators \cite{chen2016hidden,liu2019pentaquark,chen2017review,dong2017description,lebed2017heavy,guo2018hadronic,albuquerque2019qcd,yamaguchi2020heavy,guo2020threshold,brambilla2020xyz}, motivating efforts to develop more general mass formulas capable of describing these new states and predicting future ones. Within a single SU(3) multiplet, small mass splittings arise from differences in the up, down, and strange quark masses. To quantify this effect, Gell‑Mann and Okubo proposed the well‑known GMO mass formula \cite{gell2018symmetries}, which serves both as a valuable phenomenological tool and as a benchmark for testing flavor‑symmetry breaking effects.

In recent years, machine learning (ML) has made significant advances across a wide range of disciplines. In physics, ML has been used to explore theoretical frameworks \cite{wetzel2020discovering,song2021ads} and to analyze high-energy phenomena \cite{hezaveh2017fast,soma2023reconstructing,pang2018equation,chen2024machine,radovic2018machine,guest2018deep,albertsson2018machine,kim2024machine,zhou2024exploring,aarts2025physics,jiang2021deep,chen2025flavor,mansouri2026holographic,dai2026extracting,mei2025neural,xiang2022determination}. Machine learning techniques enable the extraction of physical information from high-dimensional data \cite{du2023examination, wu2026deep, li2025hep, wu2025jet}—tasks that are often difficult to accomplish using traditional methods. Neural networks, in particular, excel at approximating nonlinear mappings and uncovering hidden patterns in both experimental and simulated data. In hadron and nuclear physics, neural networks have been successfully employed to predict meson properties with spontaneously emerging symmetries~\cite{tong2026meson}, to extract the strong coupling constant across the full energy range~\cite{wang2024analysis}, to determine QCD parameters in magnetized matter~\cite{ding2026determining}, and to predict nuclear masses with quantified uncertainties~\cite{qu2025nuclear}. In addition, machine learning methods have been widely used in broader areas of physics, including studies of phase transitions and representation learning, further highlighting their capability in capturing fundamental patterns in physical systems~\cite{bai2024universal, ma2023phase,li2023machine}. Physics-informed neural networks (PINNs) \cite{raissi2019physics,chiu2022can,wen2025application} and their extensions embed physical constraints directly into the learning process, thereby bridging the gap between data-driven models and theoretical frameworks.

Among emerging neural network architectures, the Kolmogorov–Arnold Network (KAN) \cite{liu2025kan,Liu2025KANScience} stands out for its interpretability and symbolic regression capabilities. Inspired by the Kolmogorov–Arnold representation theorem, KAN replaces the fixed weights of conventional neural networks with learnable univariate functions, enabling it to represent complex multivariate relationships in a form that is both flexible and analytically tractable. By expressing learned relations in explicit symbolic form, KAN explicitly addresses the black-box limitation of conventional neural networks and enables transparent physical interpretation. Unlike traditional neural networks \cite{liu2021machine}, symbolic regression methods \cite{udrescu2020ai,schmidt2009distilling}, or genetic algorithms \cite{zhang2022approach}, KAN combines the expressive power of neural networks with the interpretability of symbolic regression, while offering higher sample efficiency and better integration of physical priors. Applications of KAN in physics are steadily expanding; for example, it has been used to model and compute the heavy-quark potential via holographic methods combined with lattice QCD data \cite{luo2025neural}. Such successes highlight KAN’s considerable potential for tackling complex scientific problems \cite{panczyk2025opening,luo2025neural,guo2025physics,hao2024first}.

In this study, we employ a KAN-based symbolic regression approach to analyze baryon mass data from SU(3) octets and decuplets. Our objectives are: (i) to recover the Gell‑Mann–Okubo (GMO) formula \cite{gell2018eightfold,okubo1962note,gell2018symmetries} using modern data‑driven techniques; (ii) to assess its generalization capability across different SU(3) multiplets; and (iii) to demonstrate how KAN can bridge machine learning and interpretable physics. Notably, we not only successfully reconstruct the classical GMO formula \cite{gell2018eightfold,okubo1962note,gell2018symmetries} but also recover underlying relationships among quantum numbers within the octet and decuplet, offering new perspectives on flavor‑symmetry breaking.

{\it Framework:} In the section, we will introduce the detailed training progress. The fundamental structure is presented below:

\noindent 1.~\textit{Input data:} In this study, we use the baryon masses of the SU(3) octet and decuplet as the primary input data. To address the limitations imposed by the extremely small sample size, we construct two complementary data construction strategies to examine the behavior of the KAN model under both idealized and realistic conditions.

In the first stage, aimed at probing the intrinsic symbolic representation capability of KAN in the absence of experimental noise, we construct an augmented (replicated) dataset by replicating the PDG central mass values for each baryon 1000 times while keeping the corresponding quantum numbers ($I$, $Y$) unchanged. This procedure corresponds to an effective zero-variance (Dirac-delta-like) limit of the data distribution, where each mass value is fixed at its central value. This procedure does not introduce additional physical information, but stabilizes the symbolic regression process under small-sample conditions.

In the second stage, to reflect realistic experimental uncertainties, we generate an additional dataset by sampling baryon masses from Gaussian distributions $\mathcal{N}(\mu, \sigma^2)$, where $\mu$ is the PDG central value and $\sigma$ corresponds to the reported experimental uncertainty. This Gaussian-perturbed dataset is constructed specifically for the decuplet analysis to evaluate the robustness of the model. Under such noise, the KAN model is still able to recover the linear GMO mass relation and accurately predict the mass of the $\Omega^-$ baryon.

The mass values are taken from the Particle Data Group (PDG) in Table~\ref{table1}~\cite{nakamura2010review}. These datasets preserve the underlying SU(3) symmetry structure while providing stable and physically meaningful inputs for the KAN neural network~\cite{liu2025kan,Liu2025KANScience} to perform symbolic regression.

Based on this dataset, we apply KAN for symbolic regression to explore the relationships between baryon masses and quantum numbers. In the decuplet analysis, we consider two complementary settings. First, using the PDG central mass values without additional perturbations, we perform a standard fit based on the linear GMO form to establish a reference description of the equal-spacing pattern.

Second, to evaluate the robustness of the model under realistic experimental conditions, we construct a Gaussian-perturbed dataset in which baryon masses are sampled according to the reported PDG uncertainties. On this dataset, KAN is applied to perform symbolic regression.

Under such noisy conditions, KAN extracts a simple linear relation between baryon masses and hypercharge, consistent with the expected equal-spacing behavior. Moreover, the learned relation enables an accurate prediction of the $\Omega^-$ baryon mass, even though it is not explicitly enforced in the fitting procedure.

These results indicate that the KAN model is able to recover the underlying GMO structure in both idealized and noise-perturbed settings, demonstrating its robustness and stability in the presence of experimental uncertainties. Building on this structure, we further perform symbolic regression to reconstruct the explicit functional form of the decuplet Gell-Mann–Okubo formula. This also provides a simple data-driven validation of the equal-spacing rule in the decuplet.

For the octet analysis, we adopt the same preprocessing and stratified sampling procedures as in the first stage for the decuplet baryons, KAN symbolic regression is then applied to investigate the dependence of baryon masses on hypercharge($Y$) and isospin($I$). The model extracts a compact and physically interpretable expression that captures the observed quadratic dependence on $Y$ and $I$. This relation serves as the basis for reconstructing the complete Gell‑Mann–Okubo formula through further symbolic regression. 
    
The final symbolic expressions obtained for both the octet and decuplet exhibit high accuracy and good interpretability, and remain consistent with theoretical expectations of SU(3) flavor-symmetry breaking \cite{gell2018eightfold,okubo1962note,gell2018symmetries}.

\noindent 2.~\textit{Model Selection:} To ensure that the symbolic regression model is both physically interpretable and capable of accurately capturing the SU(3) baryon mass patterns, we implemented a systematic model-selection process.

Initially, a custom-parameterized KAN model was trained using the L-BFGS optimizer, with sparsity-promoting regularization explicitly incorporated into the objective function. In particular, two regularization terms ($lamb$ and $lambda\_entropy$) were introduced to penalize redundant functional components and reduce structural complexity. All training procedures were conducted under fixed random seeds to ensure reproducibility.

After each training stage, a deterministic pruning operation (model.prune) was applied. This built-in procedure removes branches with negligible contributions based on the learned functional weights, thereby simplifying the network by retaining only the dominant functional pathways. Importantly, no manually defined pruning thresholds were introduced in this work. Instead, sparsity is governed entirely by the regularization terms during training, making the pruning process fully data-driven and reproducible.

Following pruning, the reduced model was retrained using the same optimization and regularization settings to refine the remaining functional structure and mitigate any potential performance degradation. This alternating procedure of regularized training, pruning, and retraining was repeated iteratively until convergence, as indicated by stable loss values and evaluation metrics.

Through this pipeline, the model progressively prunes redundant components and converges to a compact symbolic representation. While the pipeline is largely automated, limited manual intervention is applied when necessary, mainly for algebraic simplification and presentation purposes. This hybrid approach allows for systematic derivation of the results, enhancing both interpretability and methodological transparency.

Upon achieving satisfactory performance, symbolic regression \cite{schmidt2009distilling} was applied. Constrained by the Physics Filter, the regression avoided overly complex functional forms, yielding a physically interpretable expression:
\begin{equation}
M = a\,Y + b\left[I(I+1) - \frac14 Y^2\right] + c. \label{eq:1}
\end{equation}
where $a$, $b$, and $c$ are free parameters determined from the data.

\noindent 3.~\textit{SU(3)-consistent generalization:} We assess generalization in the sense of SU(3) multiplet consistency, both at the level of symbolic structure and representation-wide coverage. For the model, a nontrivial requirement is that a single invariant symbolic form be shared by all members of a given multiplet—for instance, the KAN-based regression of the Gell–Mann–Okubo relation for octet baryons retains a consistent functional structure across all multiplet members. For the data, the model must perform reliably for all baryons within a given SU(3) representation. 

\noindent 4.~\textit{{Loss function:}} Model performance is evaluated using the mean squared error \cite{hastie2009elements,bishop2006pattern,goodfellow2016deep,geman1992neural,murphy2012machine,hastie2009introduction} (MSE), defined as
\begin{equation}
\text{MSE} = \frac{1}{n} \sum_{i=1}^n (\hat{y}_i - y_i)^2.
\end{equation}
This metric measures the average squared deviation between predictions and observations. During training, numerical gradient descent with a chosen learning rate minimizes the MSE, optimizing model parameters for accuracy. In symbolic regression \cite{schmidt2009distilling}, optimization extends beyond parameter fitting to include structural discovery. Thus, MSE not only quantifies prediction error but also facilitates the joint evolution of model structure and parameter convergence.

\begin{table}[h!]
\centering
\caption{Baryon multiplet properties under SU(3) flavor symmetry. Masses (in MeV) represent isospin-averaged values from PDG database \cite{nakamura2010review}, with uncertainties spanning the full range of observed states within each multiplet. Decuplet states ($J^{P}=3/2^{+}$) and octet states ($J^{P}=1/2^{+}$) are distinguished by their spin-parity quantum numbers and quark composition.} 

\label{table1}
\scalebox{0.75}{ 
\begin{tabular}{l l c c c}
\toprule[1.5pt]
\textbf{Name} & \textbf{Symbol} & \textbf{I} & \textbf{Y} & \textbf{Mass (MeV)} \\
\midrule[1pt]
\multicolumn{5}{c}{\textbf{Octet}} \\ 
\midrule[0.5pt]
Nucleons & $N$ & $\frac{1}{2}$ & 1 & $939\pm1$ \\
Lambda baryons & $\Lambda$ & 0 & 0 & $1116\pm1$ \\
Sigma baryons & $\Sigma$ & 1 & 0 & $1193\pm4$ \\
Xi baryons & $\Xi$ & $\frac{1}{2}$ & -1 & $1318\pm3$ \\
\midrule[0.5pt]
\multicolumn{5}{c}{\textbf{Decuplet}} \\
\midrule[0.5pt]
Delta baryons & $\Delta$ & $\frac{3}{2}$ & 1 & $1232\pm2$ \\
Sigma baryons & $\Sigma^*$ & 1 & 0 & $1385\pm3$ \\
Xi baryons & $\Xi^*$ & $\frac{1}{2}$ & -1 & $1533\pm2$ \\
Omega baryon & $\Omega$ & 0 & -2 & $1672\pm1$ \\
\bottomrule[1.5pt]
\end{tabular}
} 
\end{table}

\noindent 5.~\textit{Training progress of decuplet baryons:} Regarding the decuplet baryons, to explore possible latent linear patterns among the SU(3) decuplet baryons, we applied a KAN-based symbolic regression to their experimental mass data. Rather than prescribing a predefined functional form, the network searched for a simple mathematical expression consistent with the data and symmetry constraints. After training, the resulting symbolic form revealed a clear linear dependence between baryon masses and their quantum numbers, consistent with the expected equal-spacing behavior within the decuplet. 

This data-driven discovery not only reaffirms the validity of the SU(3) flavor symmetry but also demonstrates the ability of the KAN model to recover fundamental linear relations directly from empirical data without prior assumptions. The extracted formula provides a quantitative bridge between the group-theoretic structure of the decuplet and the experimentally observed mass spectrum.

Figure \ref{fig:decuplet linear} illustrates the KAN neural network architecture employed to investigate the linear relations in the decuplet, which consists of two hidden layers. As shown in Table \ref{tab:decuplet linear}, the final results were obtained after two rounds of training and pruning. Beyond these two iterations, further training produced no change in the loss function, indicating that the model had converged. The final formula obtained for decuplet baryons is:
\begin{figure}[H]
    \centering
    \includegraphics[width=9cm]{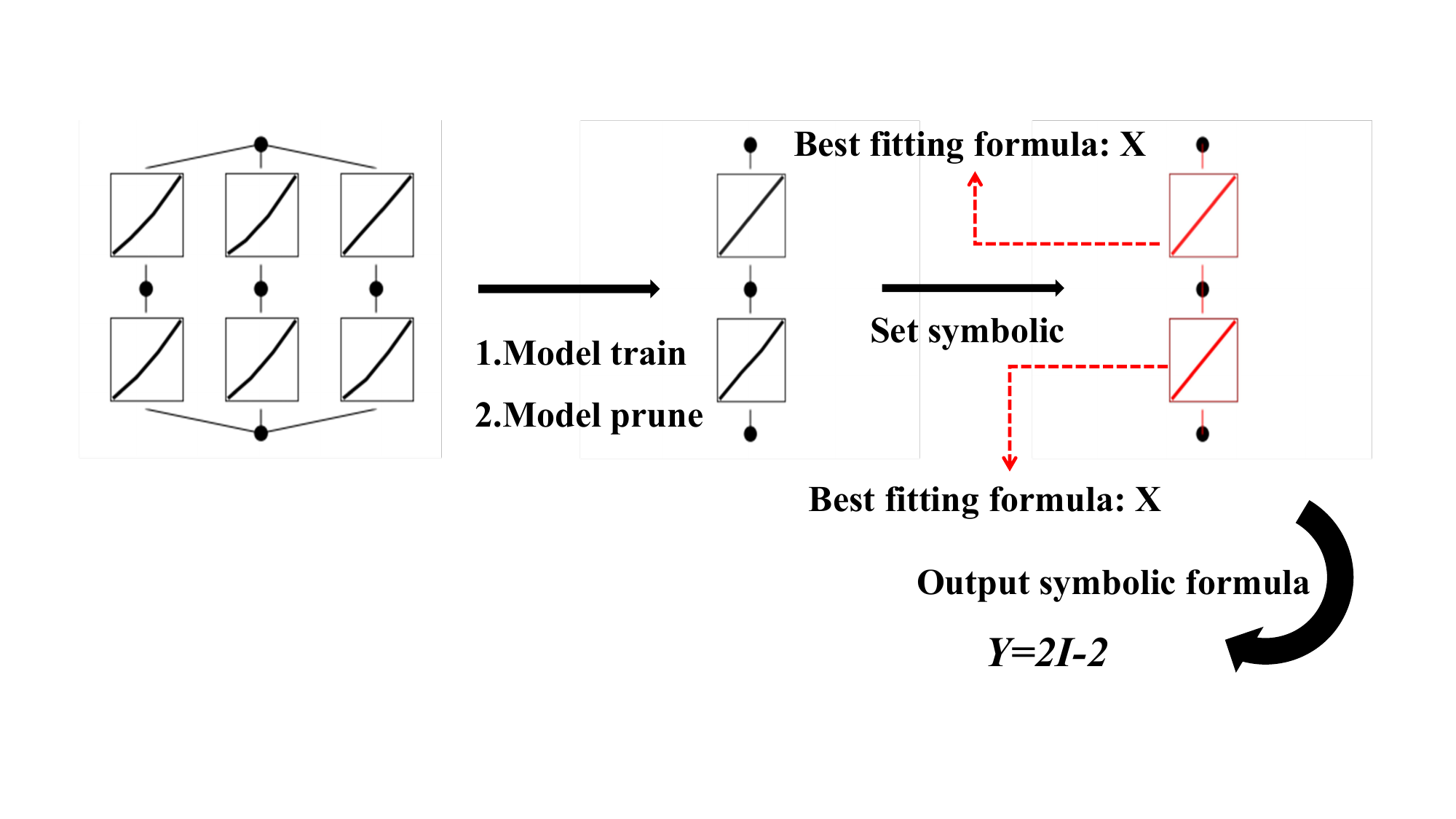}
    \caption{Figure 1 illustrates the process of using a KAN neural network to uncover linear relationships among decuplet baryons.}
    \label{fig:decuplet linear}
\end{figure}
\begin{table}[htbp]
\centering
\caption{training log of decuplet linear}

\label{tab:decuplet linear}
\begin{tabular}{l S[table-format=1.2e-2, table-column-width=2.5cm] S[table-format=1.2e-2, table-column-width=2.5cm]}
\toprule[1.5pt]
\textbf{Step} & \textbf{train loss} & \textbf{test loss} \\
\midrule[1pt]
1 & 4.05e-02 & 4.29e-02 \\
2 & 3.25e-10 & 1.73e-10 \\
\bottomrule[0.5pt]
\end{tabular}
\end{table}

\begin{equation}
Y_{10} = 2I_{10} - 2.\label{eq:5}
\end{equation}

Building on the identified linear patterns within the SU(3) decuplet, we performed symbolic regression to recover the explicit functional dependence between baryon masses and their quantum numbers. Using experimental data from PDG \cite{nakamura2010review}, the KAN model was trained to learn this mapping in a data-driven yet physically constrained manner. The regression converged rapidly with low loss, indicating that the network effectively captured the underlying symmetry structure.

\begin{figure*}[htbp]
    \centering
    \includegraphics[width=0.9\linewidth]{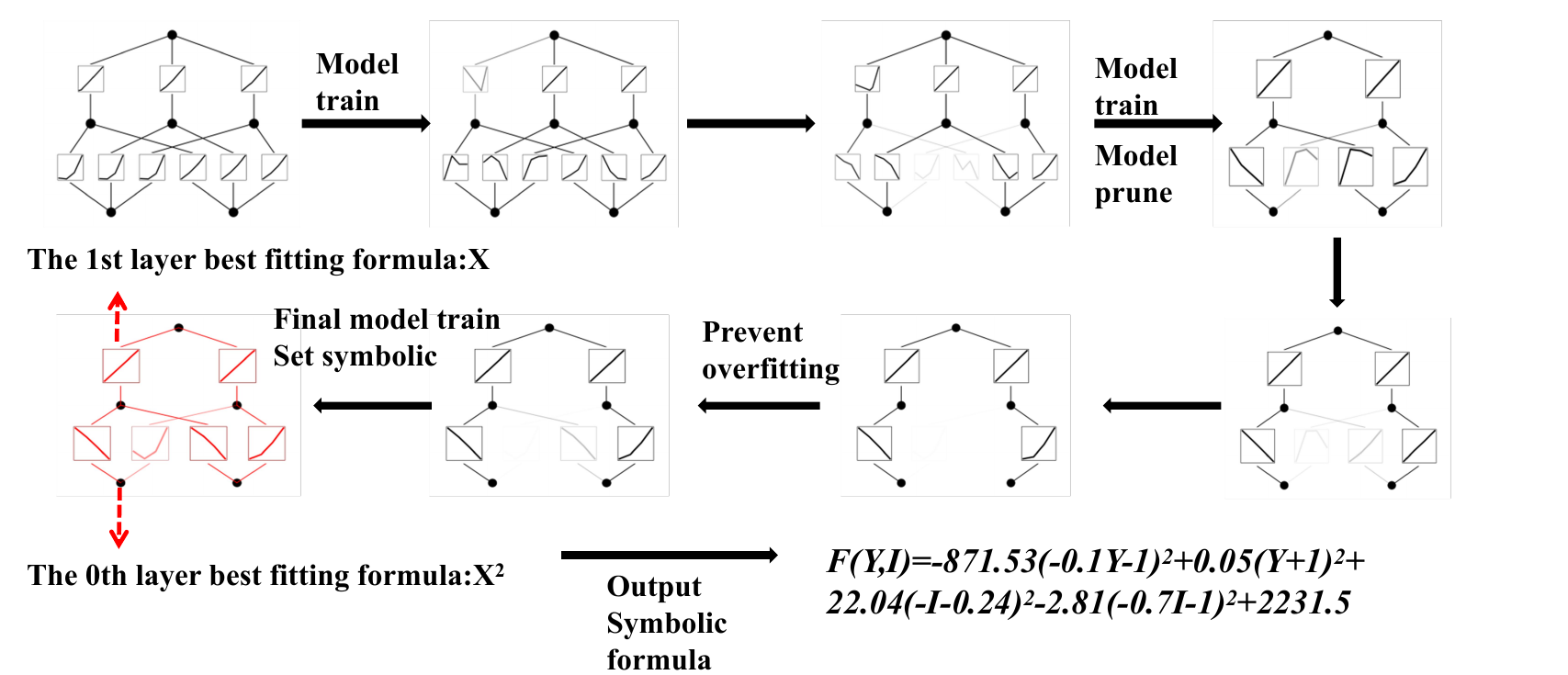}
    \caption{Figure 2 illustrates the key process of using a KAN neural network to discover the Gell-Mann–Okubo formula for decuplet baryons. including only the essential steps such as key training, pruning, and symbolic regression.} 
    \label{fig:decuplet gellmann}
\end{figure*}

Further, to obtain the GMO formula for the decuplet, we constructed a KAN with a (2,3,1) architecture, corresponding to the number of nodes in the input, hidden, and output layers, respectively, as shown in Figure \ref{fig:decuplet gellmann}. The input variables are hypercharge $Y$ and isospin $I$, and the output is the mass $M$. During training, the regularization parameters and learning rate were adjusted adaptively after each iteration to reduce the loss to an appropriate level. At the same time, the model automatically identified and removed redundant or ineffective branches, thereby improving generalization and reducing computational complexity. Through pruning, the network structure was simplified, retaining only the paths that contributed to the final symbolic regression. After ten training iterations, as shown in Table \ref{tab:decuplet gellmann}, the loss function exhibited no further change, indicating that the model had converged.


The analytical expression obtained from symbolic regression is:
\begin{multline}
F(Y_{10}, I_{10})= -871.53(-0.1Y_{10} - 1)^2 + 0.05(Y_{10} + 1)^2 \\
+ 22.04(-I_{10} - 0.24)^2 - 2.81(-0.7I_{10} - 1)^2 + 2231.5.\label{eq:6}
\end{multline}
Expanding the expression and rounding each term to two decimal places, we obtain:
\begin{multline}
F(Y_{10}, I_{10}) = -8.67Y_{10}^2 - 174.21Y_{10} + 20.66I_{10}^2 \\ 
+ 6.65I_{10} + 1358.48. \label{eq:8}
\end{multline}
According to the constraints of the GMO formula, combining with the relation $ Y_{10} = 2I_{10} - 2 $, we proceed to simplify the expression step-by-step to approximate the target form:
\begin{equation}
\begin{split}
F(Y_{10}, I_{10}) = & \; 20.66I_{10}^2 + 20.66I_{10} - 14.01I_{10} \\
              & - 8.67Y_{10}^2 - 174.21Y_{10} + 1358.48.
\end{split}
\label{eq:9}
\end{equation}
Next, extract the common factor $ Y_{10}(Y_{10} + 1) $ from the $ Y_{10} $-related terms and rewrite the remaining components:
\begin{equation}
\begin{split}
F(Y_{10}, I_{10}) = & \; 20.66I_{10}(I_{10} + 1) - 181.215Y_{10} \\
              &- 8.67Y_{10}^2 + 1344.47.
\end{split}
\label{eq:10}
\end{equation}

To achieve a more compact expression, rewrite the expression of terms involving $ Y_{10} $ and $ I_{10}(I_{10} + 1) $, and replace $F(Y_{10}, I_{10})$ with $M_{10}$ to obtain the final form, For details, see Appendix A:
\begin{equation}
\begin{split}
M_{10} = & \; -90.51Y_{10}^2 - 672.27Y_{10} + 689.73 \\
              & + 348.03I_{10}(I_{10} + 1). \\
\end{split}
\label{eq:11}
\end{equation}

This result not only validates the physical interpretability of the KAN-based regression but also highlights its potential as a tool for uncovering the underlying physical patterns encoded in the data.

\begin{table}[H]
\centering
\caption{Training log of the decuplet GMO.}
\label{tab:decuplet gellmann}
\begin{tabular}{l S[table-format=1.2e-2, table-column-width=2.5cm] S[table-format=1.2e-2, table-column-width=2.5cm]}
\toprule[1.5pt]
\textbf{Step} & \textbf{train loss} & \textbf{test loss} \\
\midrule[1pt]
1 & 1.11e-03 & 1.15e-03 \\
2 & 1.30e-03 & 1.27e-03 \\
3 & 1.51e-03 & 1.61e-03 \\
4 & 1.65e-02 & 1.73e-02 \\
5 & 2.52e-02 & 2.37e-02 \\
6 & 7.60e-04 & 7.28e-04 \\
7 & 5.02e-04 & 4.78e-04 \\
8 & 4.37e-03 & 4.20e-03 \\
9 & 4.02e-04 & 3.75e-04 \\
10 & 9.90e-04 & 9.80e-04 \\
\bottomrule[0.5pt]
\end{tabular}
\end{table}

The equal‑spacing rule is a remarkably simple yet striking consequence of the GMO mass formula for the baryon decuplet. It states that within the decuplet, each decrease of one unit in strangeness $S$(corresponding to the addition of a strange quark $s$) leads to an approximately constant increase in the particle’s mass.
We can verify this with experimental data:
\begin{enumerate}
    \item The mass difference from $\Delta$ (S=0) to $\Sigma^*$ (S=-1):
    $$ M(\Sigma^*) - M(\Delta) \approx 1385 - 1232 = 153 \; \text{MeV}. $$

    \item The mass difference from $\Sigma^*$ (S=-1) to $\Xi^*$ (S=-2):
    $$ M(\Xi^*) - M(\Sigma^*) \approx 1533 - 1385 = 148 \; \text{MeV}. $$

    \item The mass difference from $\Xi^*$ (S=-2) to $\Omega^-$ (S=-3):
    $$ M(\Omega^-) - M(\Xi^*) \approx 1672 - 1533 = 139 \; \text{MeV}. $$
\end{enumerate}

As can be seen, these three mass differences are very close, all around 140 -- 150 $\text{MeV}$. This linear, equally spaced mass relationship is the equal spacing rule of the decuplet. Consequently, the GMO formula reduces to a linear relation,
\begin{equation}
\begin{split}
M = M_0 + aY.
\end{split}
\end{equation}

Here, $M$ is the baryon mass, and $M_0$ and $a$ are constants parametrizing the SU(3) symmetric contribution and its leading symmetry breaking, respectively. Similarly, we use KAN to train the data in Table \ref{table1} \cite{nakamura2010review} to validate the equal‑spacing rule \cite{gell2018eightfold,okubo1962note,gell2018symmetries} in the baryon decuplet. We perform symbolic regression directly on the data to test whether a linear relation between baryon mass $M$ and hypercharge $Y$ can be recovered. 

For reference, we first consider the noise-free setting using only the PDG central mass values. In this idealized case, the dataset is deterministic, and KAN successfully recovers the expected linear equal-spacing relation with very small training and test errors (at the level of $10^{-4}$–$10^{-3}$), providing a baseline description of the decuplet structure.

The corresponding noise-free decuplet equal-spacing formula is:

\begin{equation}
\begin{split}
        M_{10} = 1382.16 - 146.71Y_{10}.
\end{split}
\end{equation}

We then consider a more realistic setting. To reflect experimental uncertainties, we construct an additional dataset by sampling decuplet baryon masses from Gaussian distributions centered at the PDG values, with standard deviations given by the reported experimental uncertainties.(The $\Omega^-$ baryon is excluded from the training set and used for validation.) This noise-perturbed dataset is used to evaluate the robustness of the model. The increase in the numerical loss compared to the noise-free case is expected, as the regression error reflects the variance of the introduced stochastic fluctuations.

\begin{figure}[H]
    \centering
    \includegraphics[width=10cm]{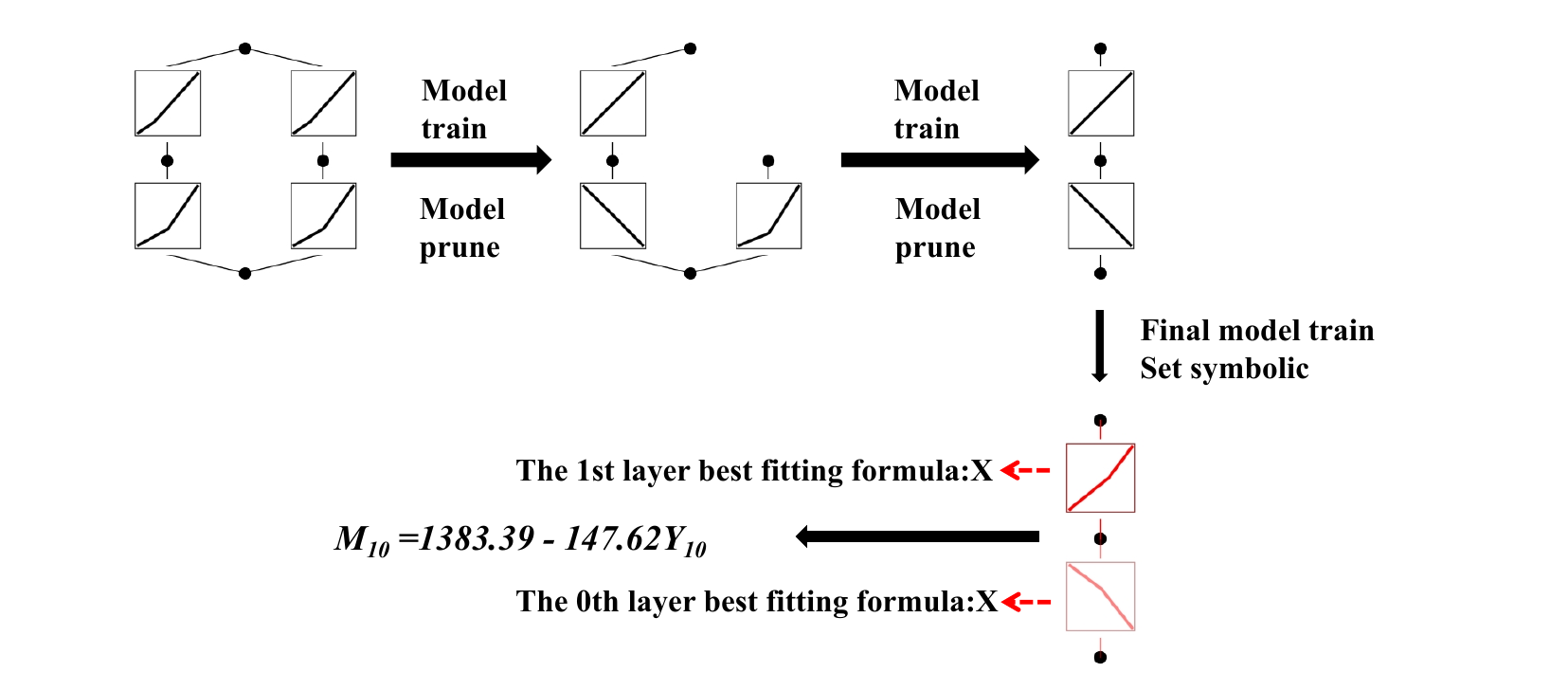}
    \caption{Figure 3 illustrates the core process of using a KAN neural network to discover the equal-spacing rule in the noise-perturbed decuplet baryons, including only the essential steps such as training, pruning, and symbolic regression.}
    \label{fig:decuplet equal}
\end{figure}

The procedure follows the same overall training workflow, including iterative training, pruning, and symbolic regression, while the specific network architecture and training configurations are adjusted to better capture the linear structure of the decuplet equal-spacing relation. Multiple training rounds are performed to ensure stable convergence, as illustrated in Fig.~\ref{fig:decuplet equal} and Table~\ref{tab:equal-spacing}. 

Under such Gaussian fluctuations, KAN-based symbolic regression still extracts a stable linear relation between baryon masses and hypercharge, consistent with the expected equal-spacing behavior of the decuplet. Moreover, the learned relation enables an accurate prediction of the $\Omega^-$ baryon mass, even though it is not explicitly enforced during training. These results demonstrate that the KAN framework maintains strong robustness and generalization capability under realistic noise conditions.

The corresponding quantitative evaluation, including RMSE, MSE and $R^2$, further confirms the high quality of the symbolic regression. The obtained noisy decuplet equal-spacing formula is given by:
\begin{equation}
\begin{split}
        M_{10} = 1383.39 - 147.62Y_{10}.
\end{split}
\end{equation}
After calculation, the root mean square error (RMSE) is approximately 4.023, the mean square error (MSE) is approximately 16.181 and the coefficient of determination $R^2$ is approximately 0.9994. 

Substituting into our decuplet equal-spacing formula, the predicted mass of the $\Omega^-$ baryon is:
\[ M(\Omega^-) \approx 1678.63 \, \text{MeV} \]
Simultaneously, we observe the spontaneous emergence of a “dominant–correction” hierarchy. Through additive composition, the model effectively decouples the fitting task: smooth activation functions capture the leading-order physical dependence, whereas irregular ones act as numerical residual compensators for local fluctuations. Although these perturbative terms lack intuitive symbolic forms, they are essential for maintaining high regression accuracy by offsetting minor errors. This synergistic mechanism demonstrates KAN’s ability to adaptively balance sparse physical representation with numerical precision.

\begin{table}[htbp]
\centering
\caption{Training log of the equal-spacing rule for the baryon decuplet}
\label{tab:equal-spacing}
\begin{tabular}{l cc}
\toprule[1.5pt]
\textbf{Step} & \textbf{train loss} & \textbf{test loss} \\
\midrule[1pt]
1 & 2.27 & 2.42 \\
2 & 2.27& 2.42 \\
\bottomrule[0.5pt]
\end{tabular}
\end{table}

\begin{figure}[H]
    \centering                  
    \includegraphics[width=1.0\linewidth]{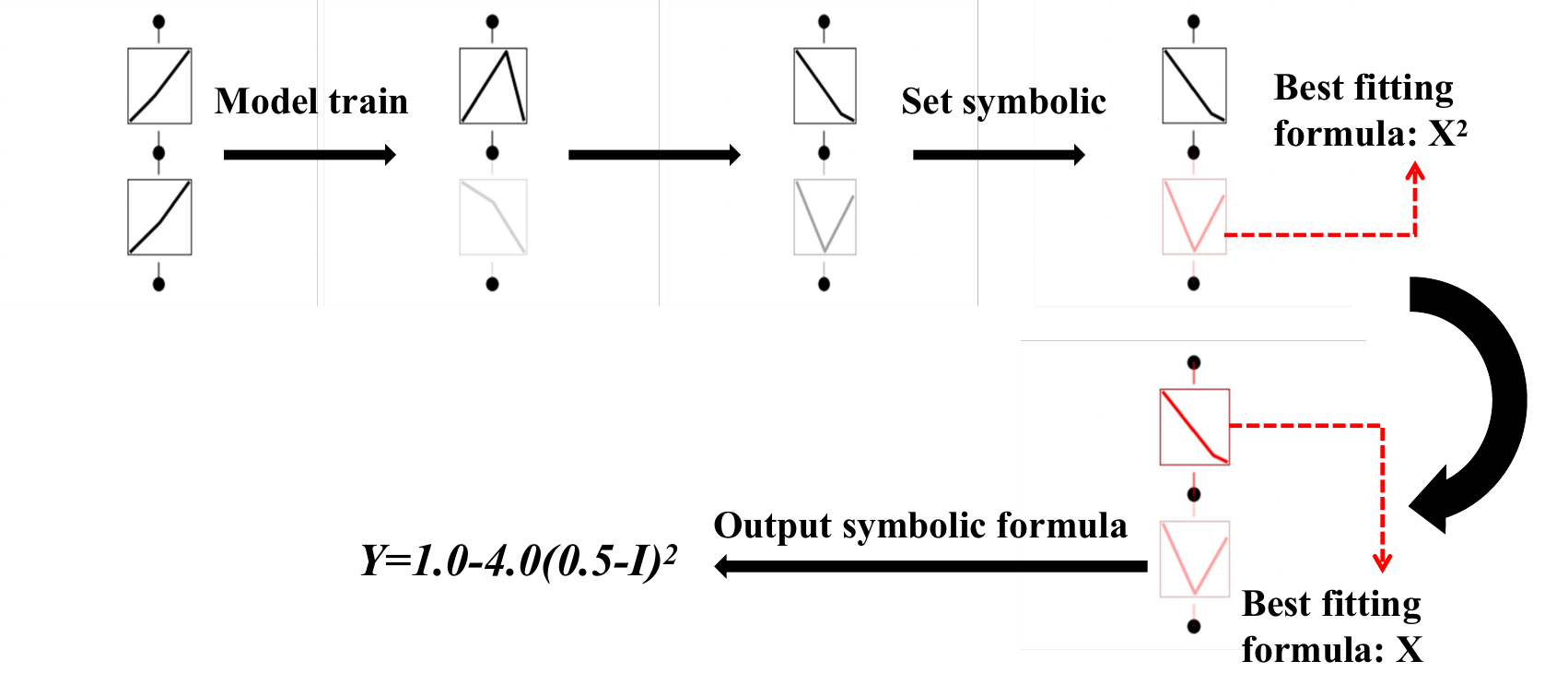}
    \caption{Figure 4 illustrates the process of using a KAN neural network to uncover binary relationships among octet baryons.}
    \label{fig:octet binary}
\end{figure}

\noindent 6.~\textit{Training progress of octet baryons:} For the octet baryons, in order to uncover their hidden nonlinear interrelationships, we performed an analysis on the SU(3) octet data using the identical preprocessing and stratified sampling procedures as in the first stage for the decuplet baryons. Through the KAN symbolic regression workflow, we focus on the relationship between hypercharge $Y$ and isospin $I$ among the octet members. Figure \ref{fig:octet binary} illustrates the KAN architecture employed for this task, which is identical to that used in deriving the linear relation for the decuplet, although the training procedure and the function fitting differ. As shown in Table \ref{tab:octet binary}, the model converged after five training iterations. Finally, symbolic regression was applied to the model, yielding the relationship between $Y$ and $I$.

\begin{equation}
Y^2_{8} = 1.0 - 4.0(0.5-I_{8})^2.\label{eq:14}
\end{equation}

\begin{table}[H]
\centering
\caption{training log of octet binary}
\label{tab:octet binary}
\begin{tabular}{l S[table-format=1.2e-2, table-column-width=2.5cm] S[table-format=1.2e-2, table-column-width=2.5cm]}
\toprule[1.5pt]
\textbf{Step} & \textbf{train loss} & \textbf{test loss} \\
\midrule[1pt]
1 & 5.35e-02 & 5.35e-02 \\
2 & 8.09e-02 & 8.04e-02 \\
3 & 9.67e-02 & 1.02e-01 \\
4 & 0.00e+00 & 0.00e+00 \\
\bottomrule[0.5pt]
\end{tabular}
\end{table}

\begin{figure*}[htbp]
    \centering
    \includegraphics[width=0.9\linewidth]{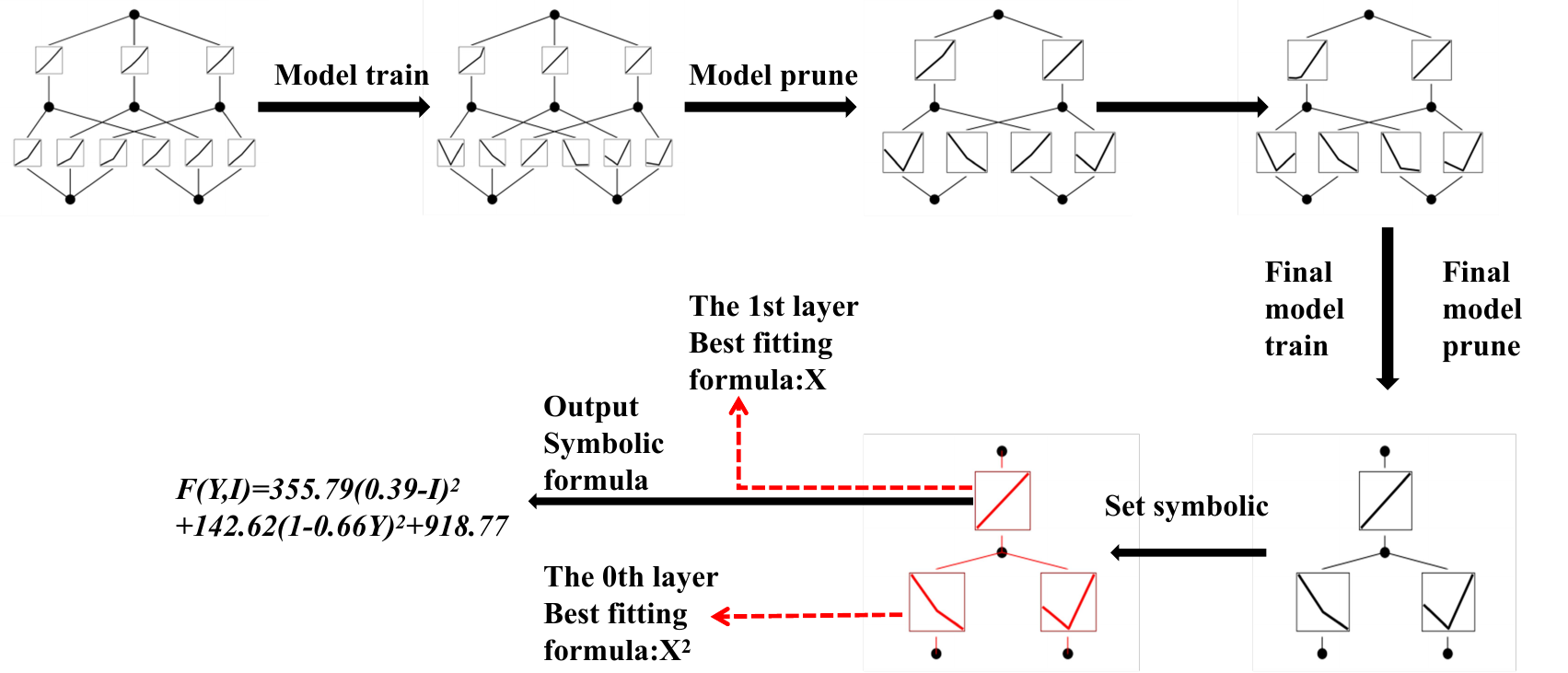}
    \caption{Figure 5 illustrates the process of using a KAN neural network to uncover the Gell-Mann–Okubo formula for octet baryons.}
    \label{fig:octet gellmann}
\end{figure*}
Then, we proceed to employ KAN to search for the GMO formula of the octet. The initial neural network architecture was identical to that used for the decuplet, but the parameter adjustments during training differed from the previous setup, with the detailed procedure shown in Figure \ref{fig:octet gellmann}. After eight training iterations, the loss function exhibited no further change, and the model was therefore considered converged, as summarized in Table \ref{tab:octet gellmann}. The figure retains only the key stages of training—such as pruning, and optimization. Finally, symbolic regression was applied to the model, yielding the initial expression of the octet GMO formula:

\begin{equation}
\begin{split}
F(Y_{8}, I_{8}) = & \;355.79(0.39 - I_{8})^2  \\
            & + 142.62(1 - 0.66Y_{8})^2 + 918.76.
\end{split}
\label{eq:15}
\end{equation}
After splitting and retaining two decimal places, simultaneously, replace $x_1$ with $Y_{8}$ and $x_2$ with $I_{8}$. the initial expression is obtained:
\begin{equation}
\begin{split}
F(Y_{8}, I_{8}) = & \;62.13Y_{8}^2 + 355.79I_{8}^2 \\
              & - 188.26Y_{8} - 277.52I_{8} + 1115.50.
\end{split}
\label{eq:1
6}
\end{equation}

To align the coefficients of $I_{8}^2$ and $I_{8}$, combine the expression above, and temporarily substitute $Y_{8}^2$ for $0.25Y_{8}^2$ (i.e., the relation $0.25Y_{8}^2 = I_{8} - I_{8}^2$), Details can be found in the Appendix A.

The $I_{8}$-related terms can be integrated into the original expression through the following transformation:
\begin{equation}
\begin{split}
& \;39.135I_{8} + 39.135I_{8}^2 - 39.135I_{8} \\
& - 39.135I_{8}^2 - 277.52I_{8} + 355.79I_{8}^2 \\
&\quad= 39.135I_{8}(I_{8} + 1) - 79.16375Y_{8}^2.
\end{split}
\label{eq:17}
\end{equation}

Substitute the above result into the original expression, consolidate all coefficients and replace $F(Y_{8}, I_{8})$ with $M_8$ to obtain the final form: 
\begin{equation}
\begin{split}
M_8 = & \; - 188.26Y_{8}  + 39.14I_{8}(I_{8} + 1) \\
          & - 17.03Y_{8}^2 + 1115.50.
\end{split}
\label{eq:18}
\end{equation}
\begin{table}[H]
\centering
\caption{training log of octet GMO}
\label{tab:octet gellmann}
\begin{tabular}{l S[table-format=1.2e-2, table-column-width=2.5cm] S[table-format=1.2e-2, table-column-width=2.5cm]}
\toprule[1.5pt]
\textbf{Step} & \textbf{train loss} & \textbf{test loss} \\
\midrule[1pt]
1 & 1.27e-02 & 1.29e-02 \\
2 & 4.60e-03 & 4.60e-03 \\
3 & 7.87e-04 & 7.85e-04 \\
4 & 8.79e-03 & 8.81e-03 \\
5 & 4.04e-03 & 4.05e-03 \\
6 & 3.64e-09 & 1.37e-09 \\
7 & 5.19e-08 & 5.32e-10 \\
8 & 2.92e-11 & 9.99e-13 \\
\bottomrule[0.5pt]
\end{tabular}
\end{table}

\noindent 7.~\textit{Output:} For the SU(3) decuplet baryons (as shown in the decuplet section of Table~\ref{tab:results}), We analyze both central-value and noise-perturbed datasets. Using the PDG central masses, the symbolic form corresponding to the GMO formula is successfully reproduced, and the linear equal-spacing relation for the decuplet is also recovered.

When Gaussian noise consistent with experimental uncertainties is introduced, KAN-based symbolic regression still extracts a stable linear relation between baryon masses and hypercharge, consistent with the GMO equal-spacing rule. Moreover, the learned relation enables an accurate prediction of the $\Omega^-$ baryon mass, demonstrating the robustness of the model under realistic noise conditions.

For the SU(3) octet baryons (as shown in the octet section of Table~\ref{tab:results}), the method reveals a multivariate correlation between $Y$ and $I$. Substituting this relation into the symbolic GMO expression yields a compact and interpretable octet GMO formula, consistent with the empirical mass hierarchy.

\begin{table}[htbp]
\centering 
\caption{The table presents all the results for the decuplet and octet baryons obtained in this study using the KAN neural network.}
\label{tab:results} 
\begin{longtable}{cc} 
\toprule
\multicolumn{1}{c}{\textbf{Decuplet}} \\
\midrule
Relation\\
\midrule
$Y_{10}=2I_{10}-2$ \\
$M_{10} = -90.51Y_{10}^2 - 672.27Y_{10} + 348.03I_{10}(I_{10} + 1) + 689.73$ \\
$M_{10}(equal-spacing)=1383.39-147.62Y_{10}$ \\
\bottomrule

\addlinespace
\toprule
\multicolumn{1}{c}{\textbf{Octet}} \\
\midrule
Relation \\
\midrule
$Y_{8}^2=1.0-4.0(0.5-I_{8})^2$ \\
$M_8=-188.26Y_{8}+39.14I_{8}(I_{8}+1)-17.03Y_{8}^2+1115.50$ \\
\bottomrule
\end{longtable}
\end{table}

\textbf{{\it Conclusion and outlook}}: In this work, we employ KAN to perform symbolic regression on experimental baryon mass data from SU(3) octets and decuplets, with the dual aims of rediscovering the GMO mass formulas and refining their functional forms directly from data. Given the extremely limited sample size, we construct two complementary data construction strategies while preserving the SU(3) structure. In the first stage, an augmented dataset is formed by replicating the PDG central mass values while keeping the corresponding quantum numbers (e.g., isospin \(I\) and hypercharge \(Y\)) fixed, which stabilizes the symbolic regression process without introducing additional physical information. In the second stage, to reflect realistic experimental conditions, an additional dataset is generated by sampling baryon masses from Gaussian distributions centered at the PDG values with standard deviations given by the reported experimental uncertainties. The mass values are taken from the PDG and related literature~\cite{gell2018eightfold,okubo1962note,gell2018symmetries,liu2022heavy,katel2024learning}. Model performance, monitored via MSE, guides the tuning of regularization and learning-rate parameters, followed by pruning to remove redundant branches and retain only the paths relevant for symbolic extraction.

Applying this workflow, KAN consistently extracts compact, physically interpretable symbolic expressions. For the octet, the model identifies a multivariate relation between \(Y\) and \(I\), which serves as the basis for reconstructing the octet GMO formula and generalizes across \(N\), \(\Lambda\), \(\Sigma\), and \(\Xi\). 

For the decuplet, we consider both idealized and noise-perturbed settings. Using the central-value dataset, the expected GMO formula is successfully recovered. When Gaussian noise consistent with experimental uncertainties is introduced, KAN is still able to extract a stable linear relation between baryon masses and hypercharge, consistent with the GMO equal-spacing rule. Moreover, the learned relation enables an accurate prediction of the \(\Omega^-\) baryon mass, demonstrating the robustness of the approach under realistic noise conditions.

Beyond validating SU(3) expectations, our results demonstrate KAN's advantage as a data-driven, assumption-light method for rediscovering empirical mass formulas. The approach suggests several promising directions, including applying KAN to more intricate hadron spectra and exotic states~\cite{garcilazo2025exotic,hyodo2025study,yang2020tetra,chen2016hidden,lyu2025evidence,collins2006regge}. The present framework can be naturally extended to more complex hadron systems, particularly to exotic hadrons, where different structural interpretations often coexist. A representative example is the $\Lambda(1405)$ resonance, whose internal structure has long been debated in terms of conventional three-quark excitations, meson-baryon molecular components, dynamically generated states, and possible two-pole structures~\cite{jido2003chiral,hyodo2012nature,bulava2024two}. Recent studies have continued to investigate this state from different perspectives, including off-shell chiral dynamics, femtoscopic correlations, SU(3) flavor filtering, production reactions, and lattice-QCD analyses~\cite{xie2026off,he2026identifying,gao2026producing,Sucunza:2026voz}.

From the perspective of interpretable machine learning, such systems provide natural testing grounds for extending the present KAN-based symbolic-regression framework beyond the reproduction of established mass relations. For exotic hadrons, different structural hypotheses---such as hadronic molecules, compact multiquark configurations, dynamically generated states, or mixed scenarios---can be encoded through distinct sets of physically motivated input variables, including threshold-related quantities, spin-coupling terms, flavor quantum numbers, channel-coupling information, pole positions, and reaction-dependent observables.

By applying KAN-based symbolic regression to these different feature representations, one can systematically compare the resulting expressions in terms of fitting accuracy, structural simplicity, parameter stability, and physical interpretability. Rather than directly determining the true underlying structure, this approach provides a transparent and unified framework for assessing which class of variables more naturally organizes the observed data. This is particularly relevant for exotic hadrons such as the $\Lambda(1405)$, where multiple competing interpretations exist and conventional model discrimination remains challenging.

This perspective highlights the broader methodological value of KAN in hadron physics, where competing physical pictures often coexist and data remain limited. Future extensions may integrate QCD inputs or expanded experimental datasets \cite{wilson1974confinement,durr2008ab,aoki2022flag,bazavov2014equation,he2024gravitational,zhang2022holographic,chen2022dynamical,li2025toward,zhou2019regressive,chen2019criticality,chen2021gluodynamics,chen2020quarkyonic,zhu2025bayesian,xing2025flavor,cao2021jet,liu2022heavy,katel2024learning,zhang2025machine,Zeng:2025tcz} to further refine symbolic forms, and improve symbolic-regression pipelines to enhance interpretability and predictive power in high-energy physics. Overall, our findings highlight KAN’s potential to bridge data-driven modeling and interpretable physical laws.

Another promising application of the KAN framework lies in the analysis of Regge trajectories\cite{winney2025regge,inopin2001hadronic,chen2024regge,collins2006regge} in hadron spectroscopy. Regge trajectories describe the empirical relation between the spin $J$ and the squared mass $M^2$ of hadrons, which is approximately linear for many hadron families.

From a methodological perspective, this problem is well suited to KAN, as the goal is not only to fit the data but also to extract an interpretable functional relation between physical observables. By applying KAN-based symbolic regression to hadron spectrum data, one can directly learn the dependence of $J$ on $M^2$ in a data-driven manner.

If the extracted relation is close to linear, this would provide an interpretable reconstruction of the conventional Regge behavior. More importantly, potential deviations from strict linearity, if identified by the model, may encode physically relevant information, such as symmetry-breaking effects, family-dependent structures, or more complex internal dynamics of hadrons.

In this sense, KAN offers a complementary tool for Regge trajectory studies, providing not only numerical fits but also explicit functional forms that can be examined for physical interpretation.

Overall, the extracted symbolic expressions exhibit high accuracy and strong interpretability, and remain consistent with theoretical expectations of SU(3) flavor-symmetry breaking. Compared with purely numerical fitting approaches, the KAN framework provides an interpretable, data-driven pathway for uncovering analytic structures and relations among hadron multiplets.

More broadly, this study demonstrates that underlying symmetry patterns can be inferred from experimental data within an interpretable machine-learning framework. In particular, the SU(3) flavor symmetry and its symmetry-breaking structure are reflected in the extracted symbolic relations without explicitly imposing group-theoretical constraints, providing a model-independent and quantitative characterization of symmetry-breaking effects associated with quark mass differences.

Importantly, the present analysis highlights the capability of KAN in small-data regimes, where only a limited number of independent hadronic states are available. Although data augmentation is used to stabilize training, it does not introduce additional physical information, and the problem remains intrinsically small-sample. Under this constraint, the successful reconstruction of the GMO relation indicates that KAN is able to extract intrinsic functional dependencies from minimal data.

It should also be noted that the symbolic expressions obtained from KAN are not unique and are typically presented in expanded polynomial forms. Their transformation into conventional representations, such as expressions involving $I(I+1)$, involves only straightforward algebraic rearrangement without introducing additional physical assumptions. This suggests that KAN identifies the underlying functional structure, while the final compact form corresponds to a physically motivated choice of basis.

From this perspective, KAN provides a systematic and interpretable framework for uncovering functional relations in hadron mass spectra. Although the current dataset does not yet allow for reliable extraction of higher-order corrections, the flexibility of the framework suggests strong potential for extension to more complex systems, where richer data may enable the identification of additional symmetry-breaking patterns in a data-driven and physically interpretable manner.

Our implementation code has been uploaded to the GitHub:\url{https://github.com/yaoyaoer0406/yaoyaoer}. We utilized version 1.0 of the KAN framework.

\section*{Acknowledgments}
This work is supported  by the National Natural Science Foundation of China (NSFC) Grant Nos: 12405154, and the European Union -- Next Generation EU through the research grant number P2022Z4P4B ``SOPHYA - Sustainable Optimised PHYsics Algorithms: fundamental physics to build an advanced society'' under the program PRIN 2022 PNRR of the Italian Ministero dell'Universit\`a e Ricerca (MUR).

\section*{References}
\bibliography{Reference}

\appendix
\section{Derivation of the GMO Relations from KAN Outputs}


\textbf{{\it Octet Baryons}}: This Appendix provides detailed algebraic steps for transforming the raw symbolic expressions obtained by KAN into the standard GMO form. The following derivation involves only straightforward algebraic manipulation. The raw symbolic expression obtained from the KAN model for the baryon octet is:
\begin{equation}
M_8 = 355.79(0.39 - I_8)^2 + 142.62(1 - 0.66Y_8)^2 + 918.76.
\end{equation}

Expanding all terms and combining like terms (while keeping two decimal places), we obtain:
\begin{equation}
M_8 = 62.13Y_8^2 + 355.79I_8^2 - 188.26Y_8 - 277.52I_8 + 1115.50.
\end{equation}

Using the implicit relation:
\begin{equation}
0.25Y_8^2 = I_8 - I_8^2,
\end{equation}
we aim to recover the standard $I(I+1)$ structure. To match the coefficients of $I_8$ and $I_8^2$, we compute:
\begin{equation}
\frac{355.79 - 277.52}{2} = 39.135.
\end{equation}

Rewriting the expression:
\begin{align}
&39.135I_8 + 39.135I_8^2 - 39.135I_8 - 39.135I_8^2 \notag\\
&\quad - 277.52I_8 + 355.79I_8^2 \\
=\, &39.135I_8(I_8 + 1) - 316.655I_8 + 316.655I_8^2 \\
=\, &39.135I_8(I_8 + 1) - 316.655 \times 0.25Y_8^2 \\
=\, &39.135I_8(I_8 + 1) - 79.16375Y_8^2.
\end{align}

Substituting back into the original expression, we obtain:
\begin{equation}
\begin{split}
    M_8 = &\; 39.135I_8(I_8 + 1) - 79.16375Y_8^2 - 188.26Y_8 \\
          &+ 62.13Y_8^2 + 1115.50.
\end{split}
\end{equation}

After rounding to two decimal places:
\begin{equation}
M_8 = 39.14I_8(I_8 + 1) - 17.03Y_8^2 - 188.26Y_8 + 1115.50.
\end{equation}


\textbf{{\it Decuplet Baryons}}: The raw symbolic expression obtained from the KAN model for the baryon decuplet is (defaults to two decimal places):
\begin{align}
M_{10} 
&= -871.53(-0.1Y_{10} - 1)^2 
+ 0.05(Y_{10} + 1)^2 \notag \\
&\quad + 22.04(-I_{10} - 0.24)^2 
- 2.81(-0.7I_{10} - 1)^2 \notag \\
&\quad + 2231.5
\end{align}
After expanding all terms and combining like terms (keeping two decimal places), we obtain:
\begin{equation}
M_{10} = -8.67Y_{10}^2 - 174.21Y_{10} + 20.66I_{10}^2 + 6.65I_{10} + 1358.48.
\end{equation}

Using the implicit relation:
\begin{equation}
Y_{10} = 2I_{10} - 2,
\end{equation}
we again aim to recover the $I(I+1)$ structure. Rewriting:
\begin{equation}
\begin{split}
    M_{10} = &\; 20.66I_{10}^2 + 20.66I_{10} - 14.01I_{10} - 8.67Y_{10}^2 \\
             &- 174.21Y_{10} + 1358.48.
\end{split}
\end{equation}

To obtain a form consistent with the GMO structure, we further rewrite the expression of $Y_{10}^2$ and $I_{10}(I_{10}+1)$. Using:
\begin{equation}
Y_{10}^2 = 4I_{10}^2 - 8I_{10} + 4,
\end{equation}
we rewrite the expression using algebraic identities 
to express it in a basis consistent with the GMO structure:
\begin{align}
M_{10} 
&= 20.66\,I_{10}(I_{10}+1) 
+ 327.37\,I_{10}(I_{10}+1) \notag \\
&\quad - 327.37\,I_{10}(I_{10}+1)
- 181.215\,Y_{10} 
- 8.67\,Y_{10}^2 \notag \\
&\quad+ 1344.47 
\tag{A16} \\[6pt]
&= 348.03\,I_{10}(I_{10}+1) 
- 327.37 \times \frac{Y_{10}^2 + 6Y_{10} + 8}{4} \notag \\
&\quad + 1344.47 
- 181.215\,Y_{10} 
- 8.67\,Y_{10}^2 
\tag{A17} \\[6pt]
&= -90.5125\,Y_{10}^2 
- 672.27\,Y_{10} 
+ 348.03\,I_{10}(I_{10}+1) \notag \\
&\quad+ 689.73 
\tag{A18} \\[6pt]
&= -90.51\,Y_{10}^2 
- 672.27\,Y_{10} 
+ 348.03\,I_{10}(I_{10}+1) \notag \\
&\quad+ 689.73 
\tag{A19}
\end{align}

\section{Comparison of KAN with other machine learning methods}
To further evaluate the performance of the KAN-based symbolic regression approach, we compare it with traditional least squares (LS), PySR (a genetic programming-based symbolic regression method), and discuss its differences to AI Feynman.
\subsection*{B.1 Comparison with Least Squares}
We first compare the KAN method with the least squares (LS) approach in terms of parameter extraction accuracy and stability. For the baryon octet, we adopt the standard GMO functional form
\begin{equation}
M = a\,Y + b\left[I(I+1) - \frac14 Y^2\right] + c,
\end{equation}
and perform LS fitting using the PDG central values.

The LS fitting yields
\begin{equation}
M_8 = -189.50\,Y + 40.00\left[I(I+1) - \frac14 Y^2\right] + 1111.50,
\end{equation}
with MSE = 10.12, RMSE = 3.18 MeV, MAE = 2.99 MeV, and $R^2 = 0.99$.

\begin{figure*}[htbp]
    \centering
    \includegraphics[width=0.75\linewidth]{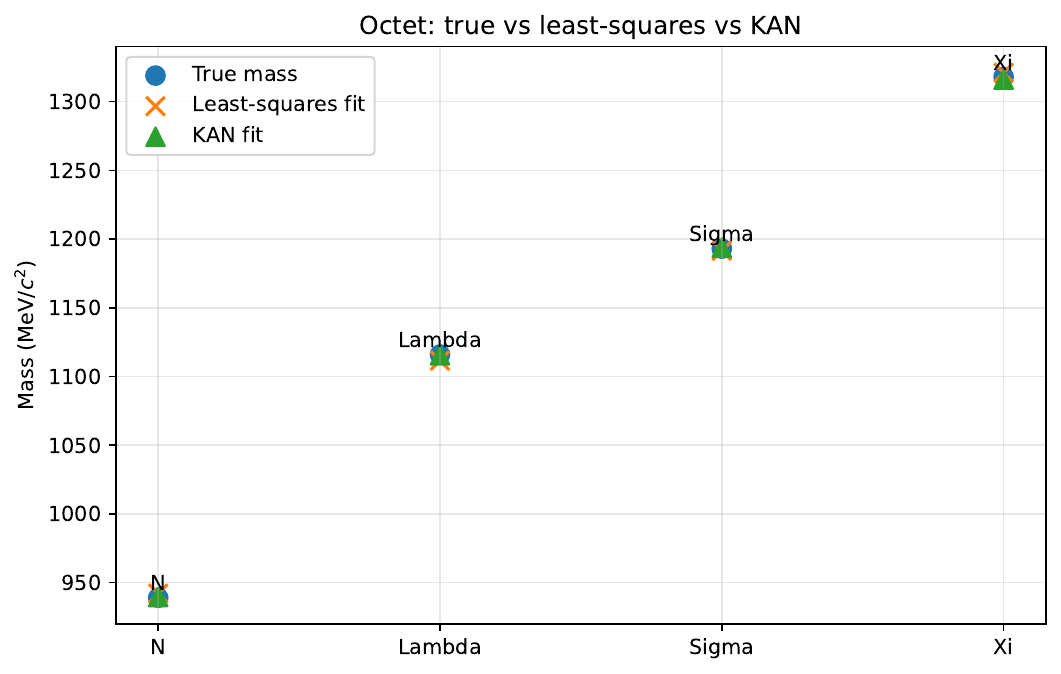}
    \caption{As shown in the Figure 6, both the least-squares method and KAN effectively fit the central values of the baryon octet, yielding good results in both cases. }
    \label{fig:octet_ls_kan_fit}
\end{figure*}

\begin{figure*}[htbp]
    \centering
    \includegraphics[width=0.75\linewidth]{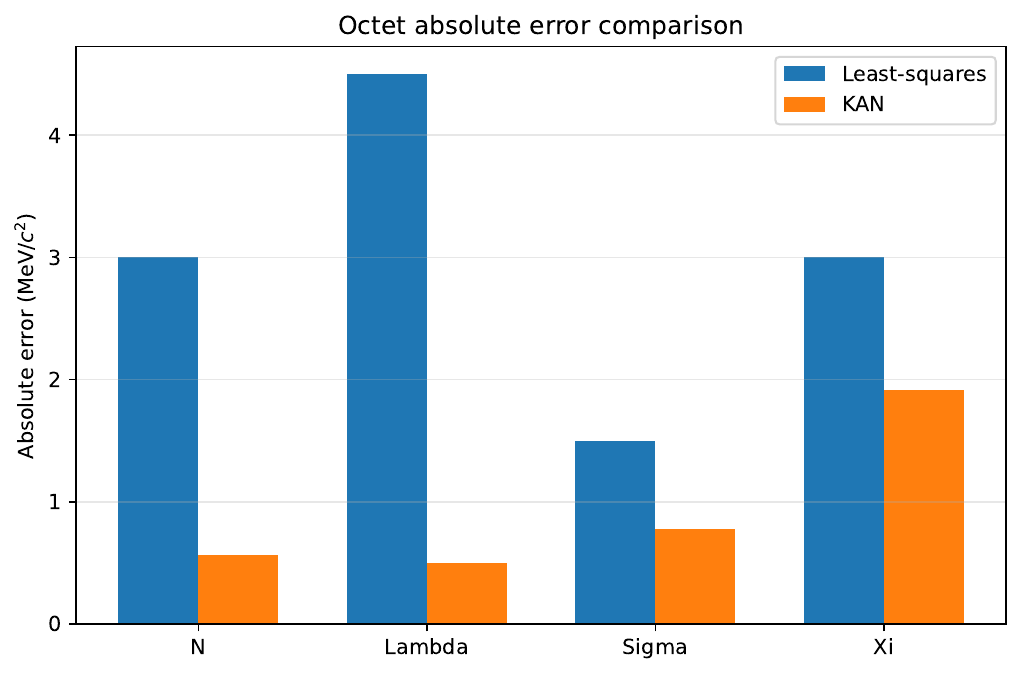}
    \caption{As shown in the Figure 7, regarding the absolute error, KAN exhibits smaller errors in the fitting of most octet baryons. This indicates that the results yielded by KAN are more closely aligned with the central values of the baryon octet. }
    \label{fig:octet_abs_error_compare}
\end{figure*}

In comparison, the KAN model learns a symbolic expression
\begin{equation}
M_{8} = -188.26Y + 39.14 I(I+1) - 17.03Y^2 + 1115.50,
\end{equation}
with improved accuracy: MSE = 1.21, RMSE = 1.10 MeV, MAE = 0.94 MeV, and $R^2 = 0.99$. As shown in the Figure~\ref{fig:octet_ls_kan_fit} indicate that both methods reproduce the octet masses with high precision, while KAN achieves better fitting accuracy as shown in the Figure~\ref{fig:octet_abs_error_compare}.

To further evaluate robustness, we perform a noise stability test by injecting Gaussian noise with standard deviation $\sigma = 5$ MeV into the mass data. Multiple independent trials are conducted (50 for KAN and 500 for LS). To ensure a fair comparison, the symbolic form is fixed to be identical for both methods, so that only the parameter extraction procedure differs.

\begin{figure*}[htbp]
    \centering
    \includegraphics[width=0.5\linewidth]{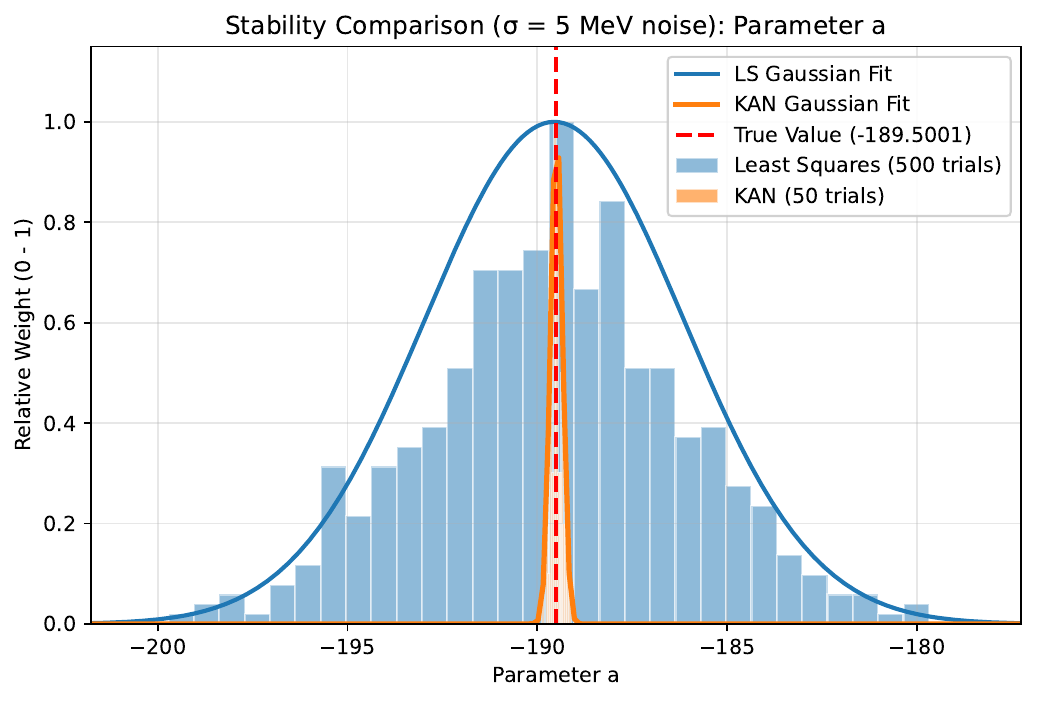}
    \caption{Figure 8 compares the fitting stability of the two methods for parameter a under Gaussian noise of $\sigma$ = 5 MeV. The figure clearly shows that, at the same noise level, the KAN method provides a more stable, less biased, and sharper distribution of parameter a than the traditional least-squares Gaussian fitting. }
    \label{fig:octet_abs_error_comparea}
\end{figure*}

\begin{figure*}[htbp]
    \centering
    \includegraphics[width=0.5\linewidth]{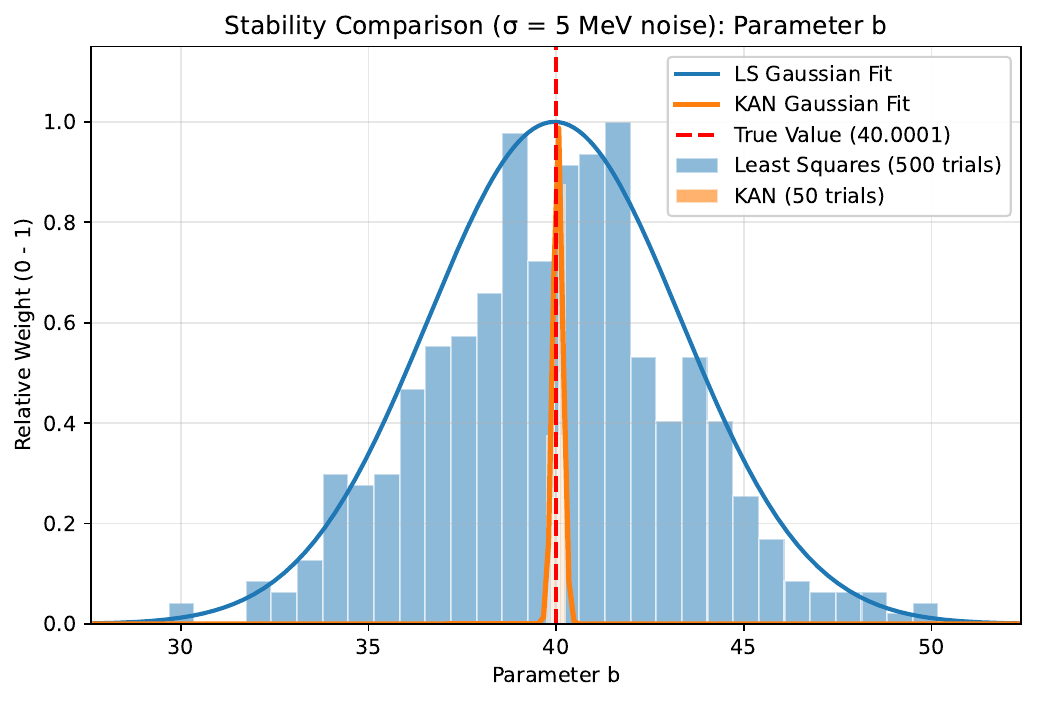}
    \caption{Figure 9 compares the fitting stability of the two methods for parameter b under Gaussian noise of $\sigma$ = 5 MeV. The figure clearly shows that, at the same noise level, the KAN method provides a more stable, less biased, and sharper distribution of parameter b than the traditional least-squares Gaussian fitting. }
    \label{fig:octet_abs_error_compareb}
\end{figure*}

\begin{figure*}[htbp]
    \centering
    \includegraphics[width=0.5\linewidth]{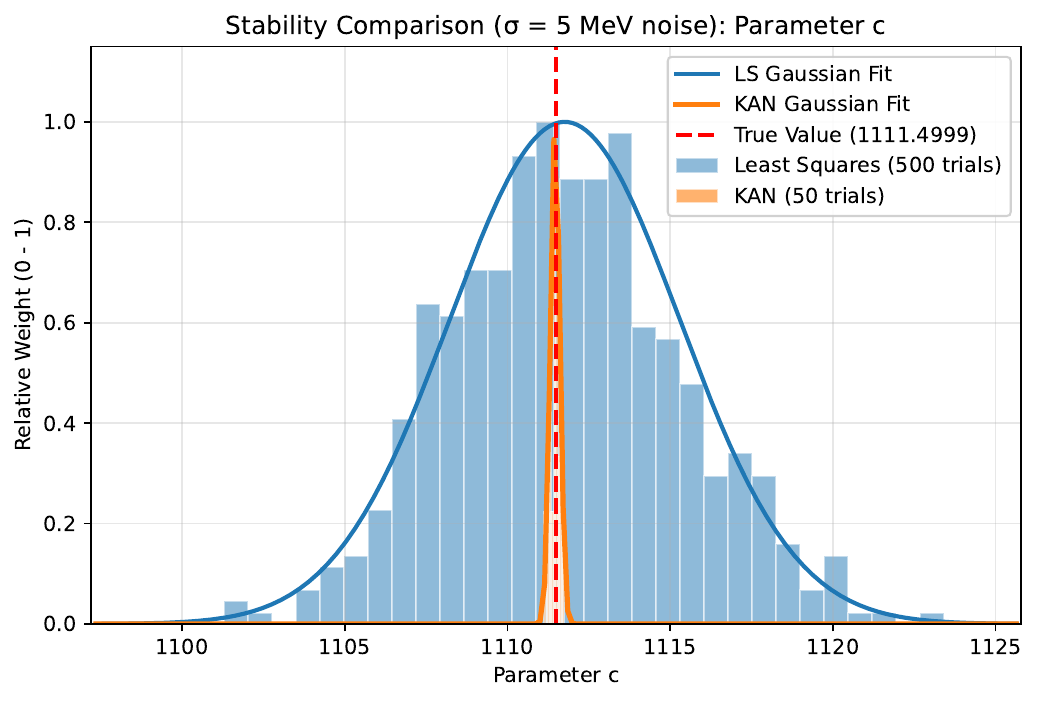}
    \caption{Figure 10 compares the fitting stability of the two methods for parameter c under Gaussian noise of $\sigma$ = 5 MeV. The figure clearly shows that, at the same noise level, the KAN method provides a more stable, less biased, and sharper distribution of parameter c than the traditional least-squares Gaussian fitting. }
    \label{fig:octet_abs_error_comparec}
\end{figure*}

We define the stability as the dispersion of the extracted parameters $(a, b, c)$ across these noisy realizations. As shown in Figures \ref{fig:octet_abs_error_comparea}, \ref{fig:octet_abs_error_compareb}, and \ref{fig:octet_abs_error_comparec}, the LS method is more sensitive to local noise fluctuations, leading to larger variations in the fitted parameters. In contrast, KAN exhibits significantly improved stability, with the extracted parameters forming a more concentrated distribution around the expected values.

This enhanced robustness can be attributed to two factors. First, the KAN training process incorporates intrinsic regularization, which suppresses overfitting to high-frequency noise. Second, KAN adopts a two-stage extraction procedure: it first captures the global functional structure using B-spline representations, and then, after fixing the symbolic form (via \texttt{fix\_symbolic}), refines the parameters within a stable functional manifold.

As a result, compared with the LS method, KAN yields parameter estimates with smaller variance under Gaussian perturbations, demonstrating its superior robustness in noisy conditions.
\subsection*{B.2 Comparison with PySR}

We further compare the KAN-based symbolic regression with PySR, a representative genetic programming-based method used in Ref.~\cite{zhang2025machine}. In that work, additional physics-inspired filtering rules are introduced to guide the search toward physically meaningful expressions.

We emphasize that no additional physics-inspired filtering or feature engineering is applied to PySR in this comparison, in order to assess its baseline performance under minimal prior assumptions.

Using only the central values of the baryon octet, KAN recovers a compact and physically interpretable expression with RMSE = 1.10 MeV. By comparison, PySR using only the basic input variables $(I, Y)$ produces the empirical expression
\begin{equation}
M_{8} = 79.999 I_{8}^2 - 189.50 Y_{8} + 1111.50,
\end{equation}
with RMSE = 3.18 MeV. Although this expression provides a reasonable numerical fit, it does not reproduce the canonical GMO structure.

We further consider a guided PySR setup by including additional features ($I^2$, $Y^2$, $I(I+1)$). In this case, the fitting accuracy improves (RMSE = 0.75 MeV); however, the resulting expressions become more complex and involve higher-order mixed terms, which deviate from the expected GMO functional form and reduce interpretability.

In addition, in our sample-efficiency test, PySR fails to recover GMO-like expressions across all tested octet subsets, whereas KAN consistently extracts compact and physically meaningful symbolic relations.

The quantitative comparison of recovered expressions and fitting accuracy is summarized in Table~\ref{tab:pysr_compare}, while the comparison of model characteristics is given in Table~\ref{tab:pysr_characteristics}. These results indicate that, under the same data and input conditions, KAN provides a better balance between fitting accuracy, structural simplicity, and physical interpretability.
\begin{table}[htbp]
\centering
\caption{Recovered expressions and fitting accuracy}
\label{tab:pysr_compare}
\begin{tabular}{lcc}
\toprule
\textbf{Method} & \textbf{Input features} & \textbf{Recovered expression} \\
\midrule

KAN & $I, Y$ 
& \makecell{$-188.26Y + 39.14 I(I+1)$ \\ $-17.03Y^2 + 1115.50$} \\

PySR raw & $I, Y$ 
& \makecell{$79.999245 I^2 - 189.49992 Y$ \\ $+ 1111.5004$} \\

PySR guided & $I, Y, I^2, Y^2, I(I+1)$ 
& \makecell{$-3.1943 I^3 Y + 19.0113 I^3$ \\ $-187.7033 Y + 1117.1418$} \\

\bottomrule
\end{tabular}

\vspace{6pt}
\small
\textbf{Note:} RMSE(KAN) = 1.1006MeV, RMSE(PySR raw) = 3.1820MeV, RMSE(PySR guided) = 0.7452MeV.
\end{table}

\begin{table}[htbp]
\centering
\caption{Model characteristics comparison}
\label{tab:pysr_characteristics}
\begin{tabular}{lccc}
\toprule
\textbf{Method} & \textbf{Complexity} & \textbf{GMO-like} & \textbf{Interpretability} \\
\midrule
KAN & 8 & Yes & High \\
PySR raw & 5 & No & Medium \\
PySR guided & 9 & No & Low \\
\bottomrule
\end{tabular}

\vspace{6pt}
\small
\textbf{Note:} 
Generally speaking, the symbolic complexity of an expression is determined collectively by the number of operational nodes, the number of terms, and the presence of nonlinear structures. This definition follows common practice in symbolic regression, where expression complexity is used as a proxy for interpretability.
\end{table}

\end{document}